\documentclass[proceedings]{stacs}
\stacsheading{2009}{277--288}{Freiburg}
\firstpageno{277}

\usepackage{amsmath, amssymb, amsthm}
\usepackage{dsfont}
\usepackage{mathrsfs}
\usepackage{algorithm}
\usepackage{algorithmic}

\theoremstyle{plain}

\newtheorem{coro}[thm]{Corollary}

\theoremstyle{definition}
\newtheorem{defn}[thm]{Definition}

\newcommand{\F}{\mathds{F}}

\newcommand{\eps}{\epsilon}

\newcommand{\eqdef}{\stackrel{\mathrm{def}}{=}}
\newcommand{\cX}{\mathcal{X}}
\newcommand{\cY}{\mathcal{Y}}
\newcommand{\cG}{G}
\newcommand{\cE}{\mathcal{E}}
\newcommand{\cL}{\mathcal{L}}
\newcommand{\cR}{\mathcal{R}}

\newcommand{\NP}{\mathsf{NP}}

\newcommand{\NNP}{\mathsf{\Sigma}_2^{\mathsf{P}}}

\newcommand{\vphi}{\varphi}
\newcommand{\pem}[1]{#1} 

\newcounter{thmMain}

\newcounter{thmMainSec}
\providecommand{\ndash}{{--}}

\begin{document}
\setlength{\pdfpageheight}{\paperheight}
\setlength{\pdfpagewidth}{\paperwidth}

\title{Almost-Uniform Sampling of Points on High-Dimensional Algebraic
  Varieties} \author[auth]{M.\ Cheraghchi}{Mahdi Cheraghchi}
\author[auth]{A.\ Shokrollahi}{Amin Shokrollahi} \address[auth]{EPFL,
  Switzerland} 
\email{{mahdi.cheraghchi, amin.shokrollahi}@epfl.ch} 
\thanks{Research supported by Swiss NSF grant 200020-115983/1.}

\keywords{Uniform Sampling, Algebraic Varieties, Randomized
  Algorithms, Computational Complexity}
\begin{abstract}
  We consider the problem of uniform sampling of points on an
  algebraic variety.  Specifically, we develop a randomized algorithm
  that, given a small set of multivariate polynomials over a
  sufficiently large finite field, produces a common zero of the
  polynomials almost uniformly at random. The statistical distance
  between the output distribution of the algorithm and the uniform
  distribution on the set of common zeros is polynomially small in the
  field size, and the running time of the algorithm is polynomial in
  the description of the polynomials and their degrees provided that
  the number of the polynomials is a constant.
\end{abstract}
\maketitle

\section{Introduction}
A natural and important class of problems in computer science deals
with random generation of objects satisfying certain properties. More
precisely, one is interested in an efficient algorithm that, given a
\emph{compact} description of a set of objects, outputs an element in
the set uniformly at random, where the exact meaning of ``compact''
depends on the specific problem in question.

Uniform sampling typically arises for problems in $\NP$. Namely, given
an instance belonging to a language in $\NP$, one aims to produce a
witness uniformly at random. Here, the requirement is stronger than
that of decision and search problems.  In a seminal paper, Jerrum,
Valiant and Vazirani \cite{ref:sampling} gave a unified framework for
this problem and showed that, for polynomial-time verifiable relations
$xRy$, uniform sampling of a witness $y$ for a given instance $x$ is
reducible to approximate counting of the witnesses, and hence, can be
efficiently accomplished using a $\NNP$ oracle. It is natural to ask
whether the requirement for an $\NNP$ oracle can be lifted. In fact,
this is the case; a result of Bellare, Goldreich, and Petrank
\cite{ref:sampling2} shows that an $\NP$ oracle is sufficient and also
necessary for uniform sampling of $\NP$ witnesses.

The $\NP$ sampling problem can be equivalently stated as follows:
Given a boolean circuit of polynomially bounded size, sample an input
that produces the output $1$ (if possible), uniformly at random among
all possibilities. This problem can be naturally generalized to models
of computation other than small boolean circuits, and an interesting
question to ask is the following: For what restricted models, the
uniform (or \pem{almost-uniform}) sampling problem is efficiently
solvable (e.g., by polynomial-time algorithms or polynomial-sized
circuits) without the need for an additional oracle? Of course if the
role of the $\NP$ oracle in \cite{ref:sampling2} can be replaced by a
weaker oracle that can be efficiently implemented, that would
immediately imply an efficient uniform sampler. While for general
$\NP$ relations the full power of an $\NP$ oracle is necessary, this
might not be the case for more restricted models.

In this work, we study the sampling problem for the restricted model
of \emph{polynomial functions}.  A polynomial function of degree $d$
over a field $\F$ (that we assume to be finite) is a mapping $f\colon
\F^n \to \F^m$ such that every coordinate of the output can be
computed by an $n$-variate polynomial of total degree at most $d$ over
$\F$. The corresponding sampling problem (that we call \emph{variety
  sampling}) is defined as follows: Given a polynomial function, find
a pre-image of a given output (that can be considered the zero vector
without loss of generality) uniformly at random.  Hence, in this
problem one seeks to sample a uniformly random point on a given
algebraic variety.  It is not difficult to show that this problem is,
in general, $\NP$-hard.  Hence, it is inevitable to relax the
generality of the problem if one hopes to obtain an efficient solution
without the need for an $\NP$ oracle.  Accordingly, we restrict
ourselves to the case where

\begin{enumerate}
\item The co-dimension of the variety (or, the number of the
  polynomials that define the variety) is \emph{small},
\item The underlying field is \emph{sufficiently large},
\item The output distribution is only required to be
  \emph{statistically close} to the uniform distribution on the
  variety.
\end{enumerate}

It is shown in \cite{ref:sampling} that almost uniform generation of
$\NP$ witnesses (with respect to the statistical distance) is possible
without using an $\NP$ oracle for self-reducible relations for which
the size of the solution space can be efficiently approximated.  The
relation underlying the variety sampling problem consists of a set of
$n$-variate polynomials over $\F$ and a point $x \in \F^n$, and it
holds if and only if $x$ is a common zero of the polynomials.
Obviously, assuming that field operations can be implemented in
polynomial time, this is a polynomial-time verifiable
relation. Moreover, the relation is self reducible, as any fixing of
one of the coordinates of the witness $x$ leads to a smaller instance
of the problem itself, defined over $n-1$ variables. Approximate
counting of the witnesses amounts to giving a sharp estimate on the
number of common zeros of the set of polynomials. Several such
estimates are available.  In particular, a result of Lang and
Weil\footnote{ This result can be seen as a consequence of the Weil
  theorem (initially conjectured in \cite{ref:Weil}) which is an
  analog of the Riemann hypothesis for curves over finite fields.}
(Theorem~\ref{thm:LangWeil}) that we will later use in the paper gives
general lower and upper bounds on the number of rational points on
varieties.  Moreover, there are algorithmic results (see
\cite{ref:AH92,ref:Adleman,ref:HI98,ref:Pila,ref:Sch85} and the
references therein) that consider the problem of counting rational
points on a given variety that belongs to a certain restricted class
of varieties over finite fields.

Thus, it appears that the result of \cite{ref:sampling} already covers
the variety sampling problem. However, this is not the case because of
the following subtleties:

\begin{enumerate}
\item Our relation is not necessarily self-reducible in the strong
  sense required by the construction of \cite{ref:sampling}. What
  required by this result is that partial fixings of the witness can
  be done in steps of at most logarithmic length (to allow for an
  efficient enumeration of all possible fixings). Namely, in our case,
  a partial fixing of $x$ amounts to choosing a particular value for
  one of the $n$ variables. The portion of $x$ corresponding to the
  variable being fixed would have length $\log q$, and in general,
  this can be much larger than $O(\log |x|)$.

\item The general Lang-Weil estimate gives interesting bounds only
  when the underlying field is fairly large.

\item The algorithmic results mentioned above, being mostly motivated
  by cryptographic or number-theoretic applications such as primality
  testing, focus on very restricted classes of varieties, for
  instance, elliptic \cite{ref:Sch85} or hyperelliptic \cite{ref:AH92}
  curves (or general plane curves \cite{ref:HI98} that are only
  defined over a constant number of variables), or low-dimensional
  Abelian varieties \cite{ref:Adleman}. Moreover, they are efficient
  in terms of the running time with respect to the logarithm of the
  field size and the dependence on the number of variables or the
  degree (whenever they are not restricted to constants) can be
  exponential.
\end{enumerate}

Hence, over large fields, \pem{fine granularity} of the self-reduction
cannot be fulfilled and over small fields, no reliable and efficient
implementation of a \pem{counting oracle} is available for our
problem, and we cannot directly apply the general sampler of
\cite{ref:sampling}.  In this work, we construct an efficient sampler
that directly utilizes the algebraic structure of the problem. The
main theorem that we prove is the following:

\setcounter{thmMain}{\value{thm}}
\setcounter{thmMainSec}{\value{section}} \newcommand{\mainTheorem}{%
  Let the integer $k > 0$ be any absolute constant, $n > k$ and $d >
  0$ be positive integers, $\eps > 0$ be an arbitrarily small
  parameter, and $q$ be a large enough prime power.  Suppose that
  $f_1, \ldots, f_k \in \F_q[x_1, \ldots, x_n]$ are polynomials, each
  of total degree at most $d$, whose set of common zeros defines an
  affine variety $V \subseteq \F_q^n$ of co-dimension $k$.  There is a
  randomized algorithm that, given the description of $f_1, \ldots,
  f_k$ and the parameter $\eps$, outputs a random point $v \in \F_q^n$
  such that the distribution of $v$ is $(6/q^{1-\eps})$-close to the
  uniform distribution on $V$.  The worst case running time of the
  algorithm is polynomial in $n, d, \log q$, and the description
  length\footnote{%
    We consider an explicit description of polynomials given by a list
    of their nonzero monomials.  } of $f_1, \ldots, f_k$.%
}
\begin{thm} {\bf (Main theorem)} \label{thm:main}
  \mainTheorem \end{thm}

Though we present the above result for affine varieties, our
techniques can be readily applied to the same problem for projective
varieties as well. At a high level, the algorithm is simple and
intuitive, and can be roughly described as follows: To sample a point
on a variety $V$ of co-dimension $k$, we first sample a
$k$-dimensional affine subspace $A$ uniformly at random and then a
random point on $V \cap A$. To make the analysis clear, we show (in
Section~\ref{sec:graph}) that the problem can be viewed as a sampling
problem on \emph{almost regular} bipartite graphs, where one can
sample a left vertex almost uniformly by picking the left neighbor of
a random edge.  The main part of the analysis
(Section~\ref{sec:variety}) is to show why this reduction holds, and
requires basic tools from Algebraic Geometry, in particular the
Lang-Weil estimate on the number of points on varieties
(Theorem~\ref{thm:LangWeil}), and details on how to deal with problems
such as varying dimension and size of the intersection $V \cap A$.
The reduction combined with the graph sampling algorithm constitutes
the sampling algorithm claimed in the main theorem.

\subsection*{Connection with Randomness Extractors}

Trevisan and Vadhan \cite{ref:TV00} introduced the notion of
\emph{samplable sources} as probability distributions that can be
sampled using small, e.g., polynomial-sized, boolean circuits. An
\emph{extractor} for samplable sources is a deterministic function
whose output, when the input is randomly chosen according to any
samplable distribution, has a distribution that is statistically close
to uniform. Assuming the existence of certain \pem{hard} functions,
they constructed such extractors.

As a natural class of samplable distributions, Dvir, Gabizon and
Wigderson \cite{ref:DGW07} considered the class of distributions that
are samplable by low-degree multivariate polynomials.  They gave a
construction of extractors for such sources over sufficiently large
finite fields that does not rely on any hardness assumption and
achieves much better parameters.  Moreover, they introduced the dual
notion of \emph{algebraic sources} that are defined as distributions
that are uniform on rational points of low-degree affine varieties,
and asked whether efficient extractors exist for such sources.  Our
main theorem shows that algebraic sources (for a wide range of
parameters) are close to samplable distributions, and hence, any
extractor for samplable distributions is also an extractor for such
algebraic sources. Very recently, Dvir \cite{ref:Zeev} gave a direct
and unconditional construction of an extractor for algebraic sources
when the field size is sufficiently large.

\section{Preliminaries and Basic Facts}

We will use a simple form of the well known Schwartz-Zippel lemma and
a theorem by Lang and Weil for bounding the number of the points on a
variety:

\begin{lem} \textrm{\bf (Schwartz-Zippel)}
  \cite{ref:Schwartz,ref:Zippel} \label{lem:SchZip} Let $f$ be a
  nonzero $n$-variate polynomial of degree $d$ defined over a finite
  field $\F_q$.  Then the number of zeros of $f$ is at most $d
  q^{n-1}$. \qed
\end{lem}

\begin{thm} \textrm{\bf (Lang-Weil)}
  \cite{ref:LangWeil} \label{thm:LangWeil} Let $n, d, r$ be positive
  integers. There exists a constant $A(n,d,r)$ depending only on
  $n,d,r$ such that for any irreducible $r$-dimensional variety $V$ of
  degree $d$ defined in a projective space $\mathds{P}^n$ over a
  finite field $\F_q$, we have $ |N-q^r| \leq (d-1)(d-2)
  q^{r-\frac{1}{2}} + A(n,d,r)q^{r-1}, $ where $N$ is the number of
  rational points of $V$ over $\F_q$. \qed
\end{thm}

This theorem can be generalized to the case of reducible varieties as
follows:

\begin{coro} \label{coro:LangWeilGeneral} Let $n, d, r$ be positive
  integers. There exists a constant $A'(n,d,r)$ depending only on
  $n,d,r$ and a constant $\delta(d)$ depending only on $d$ and integer
  $s$, $1 \leq s \leq d$, such that for any $r$-dimensional variety
  $V$ of degree $d$ defined in a projective space $\mathds{P}^n$ over
  a finite field $\F_q$ we have $ |N-s q^r| \leq \delta(d)
  q^{r-\frac{1}{2}} + A'(n,d,r)q^{r-1}, $ where $N$ is the number of
  rational points of $V$ over $\F_q$.
\end{coro}

\begin{proof}
  Let $V_1 \cup V_2 \cup \ldots \cup V_t$, where $1 \leq t \leq d$, be
  a decomposition of $V$ into distinct irreducible components and
  denote the set of $r$-dimensional components in this decomposition
  by $S$. Let $s := |S|$.  Note that each component $V_i \notin S$ has
  dimension at most $r-1$ and by Theorem~\ref{thm:LangWeil}, the
  number of points on the union of the components outside $S$ is
  negligible, namely, at most $A'' q^{r-\frac{3}{2}}$ where $A''$ is a
  parameter depending only on $n, d, r$. Hence to prove the corollary,
  it suffices to bound the number of points on the union of the
  components in $S$.

  For each component $V_i \in S$ we can apply
  Theorem~\ref{thm:LangWeil}, which implies that the number of points
  of $V_i$ in $\mathds{P}^n$, assuming that its degree is $d_i$, is
  bounded from $q^r$ by at most $ (d_i-1)(d_i-2) q^{r-\frac{1}{2}} +
  \alpha_i q^{r-1}, $ for some $\alpha_i$ that depends only on $n,
  d_i, r$.  This upper bounds the number of points of $V$ by
  \[
  \sum_{i = 1}^s |V_i| \leq s q^{r} + \delta_1 q^{r-\frac{1}{2}} + A_1
  q^{r-1},
  \]
  where $\delta_1 \eqdef \sum_{i=1}^{s} (d_i-1)(d_i-2) \leq d^2$ (from
  the fact that $\sum_{i=1}^{s} d_i \leq d$) and $A_1 \eqdef
  \sum_{i=1}^{s} \alpha_i$. Note that $A_1$ and $\delta_1$ can be
  upper bounded by quantities depending only on $n,d,r$ and $d$,
  respectively. This proves one side of the inequality.

  For the lower bound on $|V|$, we note that the summation above
  counts the points at the intersection of two irreducible components
  multiple times, and it will be sufficient to discard all such points
  and lower bound the number of points that lie on exactly one of the
  components. Take a distinct pair of the irreducible components,
  $V_i$ and $V_j$. The intersection of these varieties defines an
  $(r-1)$-dimensional variety, which by the upper bound we just
  obtained can have at most $s_{ij} q^{r-1} + \delta_2 q^{r-1.5} + A_2
  q^{r-2})$ points, for some $s_{ij} \leq d^2$, and parameters
  $\delta_2$ depending only on $d$ and $A_2$ depending on $n, k, r$.
  Hence, considering all the pairs, the number of points that lie on
  more than one of the irreducible components is no more than
  $\binom{d}{2} (d^2 q^{r-1} + \delta_2 q^{r-1.5} + A_2 q^{r-2})$,
  which means that the number of distinct points of $V$ is at least $
  \sum_{i = 1}^s |V_i| - d^4 q^{r-1} - d^2 \delta_2 q^{r-\frac{3}{2}}
  - d^2 A_2 q^{r-2}, $ which is itself at least $ sq^{r} - \delta_1
  q^{r-\frac{1}{2}} - (A_1+d^4) q^{r-1} - d^2 \delta_2
  q^{r-\frac{3}{2}} - d^2 A_2 q^{r-2}.  $ Taking (crudely) $A'(n,d,r)
  \eqdef A_1+d^2 A_2+d^4 + d^2 \delta_2 + A''$ and $\delta \eqdef
  \delta_1$ proves the corollary.
\end{proof}

\begin{remark} \label{rem:LangWeilAffine}
  Corollary~\ref{coro:LangWeilGeneral} also holds for affine
  varieties.  An affine variety $V$ can be seen as the restriction of
  a projective variety $\bar{V}$ to the affine space, where no
  irreducible component of $\bar{V}$ is fully contained in the
  hyperplane at infinity.  Then the \emph{affine dimension} of $V$
  will be the (top) dimension of $\bar{V}$, and the bound in
  Corollary~\ref{coro:LangWeilGeneral} holds for $V$ if the affine
  dimension of the variety is taken as the parameter $r$ in the
  bound. This is because each irreducible component of $\bar{V}$
  intersects the hyperplane at infinity at a variety of dimension less
  than $r$, and by Theorem~\ref{thm:LangWeil}, adding those points to
  the estimate will have a negligible effect of order
  $q^{r-\frac{3}{2}}$.
\end{remark}

Finally, we review some basic notions that we use from probability
theory.  The \emph{statistical distance} (or \emph{total variation
  distance}) of two distributions $\cX$ and $\cY$ defined on the same
finite space $S$ is defined as $ \frac{1}{2} \sum_{s \in S}
|\Pr_\cX(s) - \Pr_\cY(s)|, $ where $\Pr_\cX$ and $\Pr_\cY$ denote the
probability measures on $S$ defined by the distributions $\cX$ and
$\cY$, respectively.  Note that this is half the $\ell_1$ distance of
the two distributions when regarded as vectors of probabilities over
$S$.  It can be shown that the statistical distance of the two
distributions is at most $\eps$ if and only if for every $T \subseteq
S$, we have $|\Pr_\cX[T] - \Pr_\cY[T]| \leq \eps$. When the
statistical distance of $\cX$ and $\cY$ is at most $\eps$, we say that
$\cX$ and $\cY$ are
$\eps$-close. 
We will also use the notion of a convex combination of distributions,
defined as follows:
\begin{defn}
  Let $\cX_1, \cX_2, \ldots, \cX_k$ be probability distributions on a
  finite set $S$ and $\alpha_1, \alpha_2, \ldots, \alpha_k$ be
  nonnegative real values that sum up to $1$. Then the \emph{convex
    combination} $ \alpha_1 \cX_1 + \alpha_2 \cX_2 + \cdots + \alpha_n
  \cX_n $ is a distribution $\cX$ on $S$ given by the probability
  measure $ \Pr_\cX(x) \eqdef \sum_{i=1}^k \alpha_i \Pr_{\cX_i}(x), $
  for $x \in S$.
\end{defn}

There is a simple connection between convex combinations and distance
of distributions:

\begin{prop} \label{prop:ConvexClose} Let $\cX, \cY,$ and $\cE$ be
  probability distributions on a finite set $S$ such that for some $0
  \leq \eps \leq 1$, $\cX = (1-\eps) \cY + \eps \cE$.  Then $\cX$ is
  $\eps$-close to $\cY$. \qed
\end{prop}

\section{A Vertex Sampling Problem} \label{sec:graph}

In this section we introduce a sampling problem on graphs, and develop
an algorithm to solve it. We will later use this algorithm as a basic
component in our construction of samplers for varieties. The problem
is as follows:

\newcommand{\rsamp}{\mathsf{RSamp}} \newcommand{\rnei}{\mathsf{RNei}}
\newcommand{\bisamp}{\mathsf{BipartiteSample}}
\begin{problem}
  \label{prob:BiSamp}
  Let $\cG$ be a bipartite graph defined on a set $\cL$ of left
  vertices and $\cR$ of right vertices. Suppose that the degree of
  every vertex on the right is between $1$ and $d$, for some $d > 1$,
  and the degree of every vertex on the left differs from an integer
  $\ell$ by at most $\delta \ell$.  We are given an oracle
  $\rsamp(\cG)$ that returns an element of $\cR$ chosen uniformly at
  random (and independently at each call), and an oracle $\rnei(v)$
  that returns the neighbor list of a given vertex $v \in \cR$.
  Construct an algorithm that outputs a random vertex in $\cL$ almost
  uniformly.
\end{problem}

Intuitively, for a bipartite graph which is regular from left and
right, sampling a vertex on the left amounts to picking a random edge
in the graph, which is in turn possible by choosing a random edge
connected to a random vertex on the right side. Here of course, the
graph is not regular, however the concentration of the left degrees
around $\ell$ allows us to treat the graph as if it were regular and
get an almost uniform distribution on $\cL$ by picking a random edge.
We will compensate the irregularity from right by using a ``trial and
error'' strategy. The pseudocode given in Algorithm~\ref{alg:BiSamp}
implements this idea. The algorithm in fact handles a more general
situation, in which a call to $\rsamp$ can fail (and return a special
failure symbol $\perp$) with some probability upper bounded by a given
parameter $p$.

\newcommand{\storeLine}[1]{\newcounter{#1}
  \setcounter{#1}{\value{ALC@line}}}
\newcommand{\refLine}[1]{\arabic{#1}}
\begin{algorithm}[!ht]
  \caption{$\bisamp$}
  \begin{algorithmic}[1]
    \REQUIRE $\cG, \rsamp, \rnei$ given as in
    Problem~\ref{prob:BiSamp}, and $p$ denoting the failure
    probability of $\rsamp$.  \medskip \STATE Let $\delta, d$ be as in
    Problem~\ref{prob:BiSamp}.  \STATE $t_0 \leftarrow \lceil
    \frac{d}{1-p} \ln (\frac{1-\delta}{\delta}) \rceil;\ $ $t
    \leftarrow t_0$ \WHILE {$t \geq 0$} \STATE $t \leftarrow t - 1;\ $
    $R \leftarrow \rsamp(\cG)$ \IF {$R \neq \perp$} \STATE $V
    \leftarrow \rnei(R)$ \STATE With probability $|V|/d$, output an
    element of $V$ uniformly at random and
    return. \storeLine{alg:output}
    \ENDIF
    \ENDWHILE
    \STATE Output an arbitrary element of $\cL$.
  \end{algorithmic}
  \label{alg:BiSamp}
\end{algorithm}

\begin{lem} \label{lem:vertexSamp} The output distribution of
  Algorithm~\ref{alg:BiSamp} is supported on $\cL$ and is
  $3\delta/(1-\delta)$-close to the uniform distribution on $\cL$.
\end{lem}

\begin{proof}
  First we focus on one iteration of the \textbf{while} loop in which
  the call to $\rsamp$ has not failed, and analyze the output
  distribution of the algorithm conditioned on the event that
  Line~\refLine{alg:output} returns a left vertex.  In this case, one
  can see the algorithm as follows: Add a special vertex $v_0$ to the
  set of left vertices $\cL$.  Bring the degree of each right vertex
  up to $d$ by connecting it to $v_0$ as many times as
  necessary. Hence, the graph $\cG$ now becomes $d$-regular from
  right.  Now the algorithm picks a random element $R \in \cR$ and a
  random neighbor of $R$ and independently repeats the process if
  $v_0$ is picked as a neighbor.

  Let $T \subseteq \cL$ be a non-empty subset of the left vertices
  (excluding $v_0$) in the graph. We want to estimate the probability
  of the event $T$.  We can write this probability as follows:
  \begin{eqnarray*}
    \Pr[T] = \sum_{r \in \cR} \Pr[T \mid R = r] \Pr[R = r] 
    = \frac{1}{|\cR|} \sum_{r \in \cR} \Pr[T \mid R=r]
    = \frac{1}{d |\cR|} \sum_{r \in \cR} |T \cap \Gamma(r)|,
  \end{eqnarray*}
  where in the last equation $\Gamma(r)$ is the set of neighbors of
  $r$ in the graph. Hence the summation can be simplified as the
  number of edges connected to $T$. This quantity is in the range $|T|
  \ell (1 \pm \delta)$, because the left degrees are all concentrated
  around $\ell$, ignoring $v_0$ which is by assumption not in
  $T$. That is,
  \begin{equation}
    \Pr[T] = \Pr[T, \lnot v_0] = \frac{|T|\ell}{d |\cR|}(1 \pm \delta), \label{eqn:prT}
  \end{equation}
  where we use the shorthand $(1 \pm \delta)$ to denote a quantity in
  the range $[1-\delta, 1+\delta]$.

  Hence the probabilities of all events that exclude $v_0$ are close
  to one another, which implies that the distribution of the outcome
  of a single iteration of the algorithm, conditioned on a
  non-failure, is close to uniform.  We will now make this statement
  more rigorous.

  The degree of $v_0$ can be estimated as
  \[ \deg(v_0) = d |\cR| - |\cL| \ell(1 \pm \delta)\] by equating the
  number of edges on the left and right side of the graph.  Similar to
  what we did for computing the probability of $T$ we can compute the
  probability of picking $v_0$ as
  \[
  \Pr(v_0) = \frac{1}{d |\cR|} \deg(v_0) = 1 - \frac{|\cL|}{d|\cR|}
  \ell(1 \pm \delta).
  \]
  Combining this with \eqref{eqn:prT} we get that
  \[
  \Pr[T \mid \lnot v_0] = \frac{\Pr[T, \lnot v_0]}{1-\Pr(v_0)} =
  \frac{|T|}{|\cL|}\left(1 \pm \frac{2\delta}{1-\delta}\right).
  \]
  Hence, the output distribution of a single iteration of the
  \textbf{while} loop, conditioned on a non-failure (i.e., the event
  that the iteration reaches Line~\refLine{alg:output} and outputs an
  element of $\cL$) is $2\delta/(1-\delta)$-close to the uniform
  distribution on $\cL$. Now denote by $\vphi$ the failure
  probability.  To obtain an upper bound on $\vphi$, note that the
  probability of sampling $v_0$ at Line~\refLine{alg:output} of the
  algorithm is at most $(d-1)/d$ since each vertex on the right has at
  least one neighbor different from $v_0$.  Hence,
  \begin{equation} \label{eqn:phi} \vphi \leq 1 - (1-p)/d
  \end{equation}
  Now we get back to the whole algorithm, and notice that if the
  \textbf{while} loop iterates for up to $t_0$ times, the output
  distribution of the algorithm can be written as a convex combination
  \[
  \mathcal{O} = (1-\vphi) \mathcal{D} + (1-\vphi)\vphi \mathcal{D} +
  \cdots + (1-\vphi)\vphi^{t_0-1} \mathcal{D} + \vphi^{t_0}
  \mathcal{E} = (1-\vphi^{t_0}) \mathcal{D} + \vphi^{t_0} \mathcal{E},
  \]
  where $\mathcal{D}$ is the output distribution of a single iteration
  conditioned on a non-failure and $\mathcal{E}$ is an arbitrary
  \emph{error distribution} corresponding to the event that the
  algorithm reaches the last line.  The coefficient of $\mathcal{E}$,
  for $t_0 \geq \frac{d}{1-p} \ln (\frac{1-\delta}{\delta})$, can be
  upper bounded using \eqref{eqn:phi} by
  \[
  \vphi^{t_0} \leq \left(1-\frac{1-p}{d}\right)^{\frac{d}{1-p} \ln
    (\frac{1-\delta}{\delta})} \leq \frac{\delta}{1-\delta}.
  \]
  This combined with the fact that $\mathcal{D}$ is
  $2\delta/(1-\delta)$-close to uniform and
  Proposition~\ref{prop:ConvexClose} implies that $\mathcal{O}$ is
  $3\delta/(1-\delta)$-close to the uniform distribution on $\cL$.
\end{proof}

\section{Sampling Rational Points on Varieties} \label{sec:variety}

Now we are ready to describe and analyze our algorithm for sampling
rational points on varieties.  For the sake of brevity, we will
present the results in this section for affine varieties. However,
they can also be shown to hold for projective varieties using similar
arguments.

We reduce the problem to the vertex sampling problem described in the
preceding section. The basic idea is to intersect the variety with
randomly chosen affine spaces in $\F_q^n$ and \pem{narrowing-down} the
problem to the points within the intersection. Accordingly, the graph
$\cG$ in the bipartite sampling problem will be defined as the
\pem{incidence} graph of the points on the variety with affine
spaces. This is captured in the following definition:

\begin{defn}
  \label{defn:incidence}
  Let $V$ be an affine variety of co-dimension $k$ in $\F_q^n$.  Then
  the \emph{affine incidence graph} of the variety is a bipartite
  graph $\cG = (L \cup R, E)$ defined as follows:
  \begin{itemize}
  \item The left vertex set is $V$,
  \item For a $k$-dimensional affine space $A$, we say that $A$
    \emph{properly intersects} $V$ if the intersection $V \cap A$ is
    non-empty and has dimension zero.  Then the right vertex set of
    $\cG$ is defined as the set of $k$-dimensional affine spaces in
    $\F_q^n$ that properly intersect $V$.
  \item There is an edge between $u \in L$ and $v \in R$ if and only
    if the affine space $v$ contains the point $u$.
  \end{itemize}
\end{defn}

Before utilizing the vertex sampling algorithm of the preceding
section, we need to develop the tools needed for showing that the
affine incidence graph satisfies the properties needed by the
algorithm. We begin with an estimate on the number of linear and
affine subspaces of a given dimension.  The estimate is
straightforward to obtain, yet we include a proof for completeness.

\begin{prop} \label{prop:subspace} Let $\F$ be a finite field of size
  $q \geq \sqrt{2k}$, and let $N_1$ and $N_2$ be the number of
  distinct $k$-dimensional linear and affine subspaces of $\F^n$,
  respectively.  Then we have
  \begin{enumerate}
  \item $|N_1/q^{k(n-k)}-1| \leq 2k/q^2,$
  \item $|N_2/q^{(k+1)(n-k)}-1| \leq 2k/q^2.$
  \end{enumerate} 
\end{prop}

\begin{proof}
  If $k=n$, then $N_1 = N_2 = 1$, and the claim is obvious. Hence,
  assume that $k < n$.  Denote by $N_{k,n}$ the number of ways to
  choose $k$ linearly independent vectors in $\F^n$. That is, $
  N_{k,n} = (q^n-1)(q^n-q)\cdots (q^n-q^{k-1}).  $ This quantity is
  upper bounded by $q^{nk}$, and lower bounded by $ (q^n-q^{k-1})^k
  \geq q^{nk}(1-k q^{k-1-n}) \geq q^{nk}(1-k / q^2).  $ Hence, the
  reciprocal of $N_{k,n}$ can be upper bounded as follows:
  \[
  \frac{1}{N_{k,n}} \leq \frac{q^{-nk}}{1-k/q^2} =
  q^{-nk}\left(1+\frac{k}{q^2} \cdot \frac{1}{1-k/q^2}\right) \leq
  q^{-nk}\left(1+\frac{2k}{q^2}\right),
  \]
  where the last inequality follows from the assumption that $q^2 \geq
  2k$.

  The number of $k$-dimensional subspaces of $\F^n$ is the number of
  ways one can choose $k$ linearly independent vectors in $\F^n$,
  divided by the number of bases a $k$-dimensional vector space can
  assume. That is, $ N_1 = N_{k,n}/N_{k,k}.  $ By the bounds above, we
  obtain
  \[
  N_1 \leq q^{nk} \cdot q^{-k^2} (1+2k/q^2) \qquad \text{and} \qquad
  N_1 \geq q^{nk} (1-k/q^2) \cdot q^{-k^2},
  \]
  which implies $|N_1/q^{k(n-k)}-1| \leq 2k/q^2$.  The second part of
  the claim follows from the observation that two translations of a
  $k$-dimensional subspace $A$ defined by vectors $u$ and $v$ coincide
  if and only if $u-v \in A$. Hence, the number of affine
  $k$-dimensional subspaces of $\F^n$ is the number of $k$-dimensional
  subspaces of $\F^n$ multiplied by the number of cosets of $A$, i.e.,
  $N_2 = N_1 q^{n-k}$.
\end{proof}

The following two propositions show that a good fraction of all
$k$-dimensional affine spaces properly intersect any affine variety of
co-dimension $k$.

\begin{prop} \label{prop:properLinear} Let $n, d, k$ be positive
  integers, and $V \subset \F_q^n$ be an affine variety of
  co-dimension $k$ defined by the zero-set of $k$ polynomials $f_1,
  \ldots, f_k \in \F_q[x_1, \ldots, x_n]$, each of degree at most
  $d$. Suppose that $v \in V$ is a fixed point of $V$. Then the
  fraction of $k$-dimensional affine spaces passing through $v$ that
  properly intersect $V$ is at least $1-B(k,n,d)/q$, where $B(n,d,k)$
  is independent of $q$ and polynomially large in $n, d, k$.
\end{prop}

\begin{proof}
  Without loss of generality, assume that $v$ is the origin, and that
  $q \geq \sqrt{2k}$.  Denote by $L$ the set of $k$-dimensional linear
  subspaces that can be parametrized as
  \[
  \left( \begin{array}{c} x_{k+1} \\ x_{k+2} \\ \vdots \\
      x_n \end{array} \right) = \left( \begin{array}{ccc}
      \alpha_{11} & \ldots & \alpha_{1k} \\
      \alpha_{21} & \ldots & \alpha_{2k} \\
      \vdots & \ddots & \vdots \\
      \alpha_{(n-k)1} & \ldots & \alpha_{(n-k)k} \\
    \end{array} \right)
  \left( \begin{array}{c} x_{1} \\ x_2 \\ \vdots \\ x_k \end{array} \right),
  \]
  where $\alpha \eqdef \{ \alpha_{11}, \ldots, \alpha_{(n-k)k} \}$ is
  a set of indeterminates in $\F_q$. Note that $|L| = q^{|\alpha|} =
  q^{k(n-k)}$, and define the polynomial ring $\cR \eqdef
  \F_q[\alpha_{11}, \ldots, \alpha_{(n-k)k}]$.  We first upper bound
  the number of \emph{bad} subspaces in $L$ whose intersections with
  $V$ have nonzero dimensions. Substituting the linear forms defining
  $x_{k+1}, \ldots, x_n$ in $f_1, \ldots, f_k$ we see that the
  intersection of $V$ and the elements of $L$ is defined by the common
  zero-set of polynomials $g_1, \ldots, g_k \in \cR[x_1, \ldots,
  x_k]$, where for each $i \in [k]$,
  \[
  g_i(x_1, \ldots, x_k) \eqdef f_i(x_1, \ldots, x_k,
  \alpha_{11}x_1+\cdots+\alpha_{1k}x_k, \ldots,
  \alpha_{(n-k)1}x_1+\cdots+\alpha_{(n-k)k}x_k).
  \]
  Each $g_i$, as a polynomial in $x_1, \ldots, x_k$, has total degree
  at most $d$ and each of its coefficients is a polynomial in
  $\alpha_{11}, \ldots, \alpha_{(n-k)k}$ of total degree at most
  $d$. Denote by $I \subseteq \cR[x_1,\ldots,x_k]$ the ideal generated
  by $g_1, \ldots g_k$.  For every $j \in [n]$, the ideal $I \cap
  \cR[x_j]$ is generated by a polynomial $h_j$. Each coefficient of
  $h_j$ can be written as a polynomial in $\cR$ with total degree at
  most $D$, where for a fixed $k$, $D$ is polynomially large in
  $d$. This can be shown using an elimination method, e.g.,
  generalized resultants or Gr\"obner bases (cf.\
  \cite{ref:CLO,ref:elimination,ref:KL}).  Take any coefficient of
  $h_j$ which is a nonzero polynomial in $\cR$.  The number of the
  choices of $\alpha$ which makes this coefficient zero is, by
  Lemma~\ref{lem:SchZip}, at most $D q^{k(n-k)-1}$. This also upper
  bounds the number of the choices of $\alpha$ that make $h_j$
  identically zero.

  A union bound shows that for all but at most $nD q^{k(n-k)-1}$
  choices of $\alpha$ none of the polynomial $h_j$ is identically
  zero, and hence the solution space of $g_1, \ldots, g_k$ is zero
  dimensional (and obviously non-empty, as we already know that it
  contains $v$). This gives an upper bound of $nD/q$ on the fraction
  of bad subspaces in $L$.

  By Proposition~\ref{prop:subspace}, the set $L$ contains at least a
  $1-2k/q^2$ fraction of all $k$-dimensional subspaces of $\F_q^n$.
  Hence, the fraction of $k$-dimensional subspaces of $\F_q^n$ that
  properly intersect $V$ is at least
  \[
  \left(1-\frac{2k}{q^2} \right) \left(1-\frac{nD}{q} \right) \geq
  \left(1-\frac{2k+nD}{q}\right).
  \]
  The claim follows by taking $B \eqdef 2k+nD$.
\end{proof}

\begin{prop} \label{prop:properAffine} Let $k, n, d$ be positive
  integers, and $V \subset \F_q^n$ be an affine variety of
  co-dimension $k$ defined by the zero-set of $k$ polynomials $f_1,
  \ldots, f_k \in \F_q[x_1, \ldots, x_n]$, each of degree at most $d$.
  The fraction of $k$-dimensional affine subspaces that properly
  intersect $V$ is at least
  \[
  d^{-k} \left( 1 - \frac{\delta(d)}{\sqrt{q}} -
    \frac{A'(n,d,n-k)+B(n,d,k)}{q} \right),
  \]
  where $\delta(\cdot), A'(\cdot), B(\cdot)$ are as in
  Corollary~\ref{coro:LangWeilGeneral} and
  Proposition~\ref{prop:properLinear}.
\end{prop}

\begin{proof}
  We use a counting argument to obtain the desired bound.  Denote by
  $N, N_1$, and $N_2$ the number of points of $V$, the number of
  $k$-dimensional subspaces and $k$-dimensional affine subspaces in
  $\F_q^n$, respectively.  Then Corollary~\ref{coro:LangWeilGeneral}
  (followed by Remark~\ref{rem:LangWeilAffine}) implies that
  \[
  N \geq s q^{n-k} - \delta(d) q^{n-k-\frac{1}{2}} -
  A'(n,d,n-k)q^{n-k-1},
  \]
  for some $s \in [d^k]$ (as the degree of $V$ is at most $d^k$).

  By Proposition~\ref{prop:properLinear}, for every $v \in V$, at
  least $N_1(1-B(n,d,k)/q)$ affine subspace pass through $v$ and
  properly intersect $V$. Hence in total $N \cdot N_1(1-B(n,d,k)/q)$
  affine spaces properly intersect $V$, where we have counted every
  such affine space at most $d^k$ times (This is because the
  intersection of $V$ and an affine space $A$ that properly intersects
  it is of size at most $d^k$, and $A$ is counted once for each point
  at the intersection). Thus, the fraction of distinct affine
  subspaces that properly intersect $V$ is at least
  \[
  \frac{N N_1 (1-B(n,d,k)/q)}{d^k N_2}.
  \]

  By the fact that $N_2 = N_1 q^{n-k}$ and the lower bound on $N$, we
  conclude that this fraction is at least
  \[
  d^{-k} \left( s - \frac{\delta(d)}{\sqrt{q}} - \frac{A'(n,d,n-k)}{q}
  \right) \left( 1- \frac{B(n,d,k)}{q} \right).
  \]
  As $s \geq 1$, this proves the claim.
\end{proof}

Now having the above tools available, we are ready to give the
reduction from variety sampling to the vertex sampling problem
introduced in the preceding section and prove our main theorem:

\vspace{3mm}
\noindent{\bf Proof of Theorem~\ref{thm:main}. }
Let $G=(L \cup R, E)$ be the affine incidence graph of $V$.  We will
use Algorithm~\ref{alg:BiSamp} on $G$. To show that the algorithm
works, first we need to implement the oracles $\rsamp$ and $\rnei$
that are needed by the algorithm.

The function $\rsamp$ simply samples a $k$-dimensional affine space of
$\F_q^n$ uniformly at random, and checks whether the outcome $A$
properly intersects $V$. To do so, one can parametrize the affine
subspace as in the proof of Proposition~\ref{prop:properLinear} and
substitute the parametrization in $f_1, \ldots, f_k$ to obtain a
system of $k$ polynomial equations in $k$ unknowns, each of degree at
most $D$ which is polynomially large in $d$.  As $k$ is an absolute
constant, it is possible to solve this system in polynomial time using
multipolynomial resultants or the Gr\"obner bases method combined with
backward substitutions. If at any point, the elimination of all but
any of the variables gives the zero polynomial, it turns out that the
system does not define a zero-dimensional variety and hence, $A$ does
not properly intersect $V$.  Also, if the elimination results in a
univariate polynomial that does not have a solution in $\F_q$, the
intersection becomes empty, again implying that $A$ does not properly
intersect $V$. In both cases $\rsamp$ fails, and otherwise, it outputs
$A$.  Furthermore, if the intersection is proper, the elimination
method gives the list of up to $D^k$ points at the intersection, which
one can use to construct the oracle $\rnei$.

Now we need to show that the graph $G$ satisfies the conditions
required by Lemma~\ref{lem:vertexSamp}.  By the argument above, the
degree of every right vertex in $G$ is at least $1$ and at most $D^k$,
which is polynomially large in $d$. Let $p$ denote the failure
probability of $\rsamp$. Then Proposition~\ref{prop:properAffine}
implies that $p \leq d^{-k}/2$ when $q \geq \max\{16 \delta^2(d),
4(A'(n,d,n-k)+B(n,d,k))\}$.

To bound the left degrees of the graph, note that each left node,
which is a point on $V$, is connected to all $k$-dimensional affine
subspaces that properly intersect $V$ and pass through the point.  The
number of such spaces is, by Proposition~\ref{prop:subspace}, at most
$q^{k(n-k)}(1+2k/q^2)$ (assuming $q \geq \sqrt{2k}$), and by
combination of Proposition~\ref{prop:subspace} and
Proposition~\ref{prop:properLinear}, at least
\[
q^{k(n-k)}\left(1-\frac{2k}{q^2}\right)\left(1-\frac{B(k,n,d)}{q}\right)
\geq q^{k(n-k)}\left(1-\frac{2k+B(k,n,d)}{q}\right).
\]
Now if we choose $q \geq (2k+B(k,n,d))^{1/\eps}$, the left degrees
become concentrated in the range $q^{k(n-k)}(1 \pm 1/q^{1-\eps})$.

Putting everything together, now we can apply
Lemma~\ref{lem:vertexSamp} to conclude that the output distribution of
the algorithm is $(6/q^{1-\eps})$-close to the uniform distribution on
$V$.

To show the efficiency of the algorithm, first note that
Algorithm~\ref{alg:BiSamp} calls each of the oracles $\rsamp$ and
$\rnei$ at most
\[
\frac{D^k}{1-p} \ln \left(\frac{1-q^{\eps-1}}{q^{\eps-1}} \right) \leq
2 D^k (1-\eps)\ln q
\]
times, which is upper bounded by a polynomial in $d, \ln q$. Hence it
remains to show that the implementation of the two oracles are
efficient. The main computational cost of these functions is related
to the problem of deciding whether a system of $k$ polynomial
equations of bounded degree in $k$ unknowns has a zero dimensional
solution space, and if so, computing the list of at most $D^k$
solutions of the system.  As in our case $k$ is a fixed constant,
elimination methods can be efficiently applied to reduce the problem
to that of finding the zero-set of a single uni-variate polynomial of
bounded degree.  A randomized algorithm is given in \cite{ref:Rabin}
for this problem that runs in expected polynomial time. Thus, we can
use this algorithm as a sub-routine in $\rsamp$ and $\rnei$ to get a
sampling algorithm that runs in expected polynomial time.  Then it is
possible to get a worst-case polynomial time algorithm by using a
\emph{time-out} trick, i.e., if the running time of the sampler
exceeds a (polynomially large) threshold, it is forced to terminate
and output an arbitrary point in $\F_q^n$. The error caused by this
can increase the distance between the output distribution of the
sampler and the uniform distribution on $V$ by a negligible amount
that can be made arbitrarily small (and in particular, smaller than
$1/q^{1-\eps}$), and hence, is of little importance.

Finally, we need an efficient implementation of the field operations
over $\F_q$. This is again possible using the algorithm given in
\cite{ref:Rabin}. Moreover, when the characteristic of the field is
small, deterministic polynomial time algorithms are known for this
problem \cite{ref:Shoup}. \qed

\section{Concluding Remarks}

We showed the correctness and the efficiency of our sampling algorithm
for varieties of constant co-dimension over large fields. Though our
result covers important special cases such as sampling random roots of
multivariate polynomials, relaxing either of these requirements is an
interesting problem. In particular, it remains an interesting problem
to design samplers that work for super-constant (and even more
ambitiously, linear in $n$) co-dimensions (in our result, the
dependence of the running time on the co-dimension is exponential and
thus, we require constant co-dimensions). Moreover, in this work we
did not attempt to optimize or obtain concrete bounds on the required
field size, which is another interesting problem. Finally, the error
of our sampler (i.e., the distance of the output distribution from the
uniform distribution on the variety) depends on the field size, and it
would be interesting to bring down the error to an arbitrary parameter
that is given to the algorithm.

\section*{Acknowledgment}

We would like to thank Fr\'ed\'eric Didier for fruitful discussions.

\end{document}